# Interrelation between Partial Coherence and Quantum Correlations

Sunho Kim,[1,*] Longsuo Li,[1,†] Asutosh Kumar,[2,‡] and Junde Wu[3,§]

[1]*Department of Mathematics, Harbin Institute of Technology, Harbin 150001, People's Republic of China*
[2]*P.G. Department of Physics, Gaya College, Magadh University, Rampur, Gaya 823001, India*
[3]*School of Mathematical Sciences, Zhejiang University, Hangzhou 310027, People's Republic of China*

Both coherence and entanglement stem from the superposition principle, capture quantumness of a physical system, and play a central role in quantum physics. In a multipartite quantum system, coherence and quantum correlations are closely connected. In particular, it has been established that quantum coherence of a bipartite state is an important resource for its conversion to entanglement [A. Streltsov *et al.*, Phys. Rev. Lett. 115, 020403 (2015)] and to quantum discord [J. Ma *et al.*, Phys. Rev. Lett. 116, 160407 (2016)]. We show here that there is a very close association between partial coherence introduced by Luo and Sun [S. Luo and Y. Sun, Phys. Rev. A 96, 022136 (2017)] and quantum correlations (quantified by quantum discord) in both directions. Furthermore, we propose families of coherence measures in terms of quantum correlations and quantum Fisher information.

## I. INTRODUCTION

The superposition principle is the most fundamental feature of quantum physics, and is the essence of quantum correlations [1–3] including entanglement. Quantum coherence is another fundamental aspect of quantum physics that stems from the superposition principle. Both coherence and entanglement capture quantumness of a physical system. However, coherence differs from entanglement in at least two aspects. Unlike entanglement, coherence is a basis-dependent quantity and is defined for single as well as multipartite systems. It is identified by the presence of off-diagonal terms in the density matrix. Two important observations with regard to basis-dependence are: (i) A quantum system can have nonvanishing quantum coherence in one basis and zero coherence in another basis. By definition, a quantum system has zero coherence if its density matrix is diagonal in the chosen basis. (ii) Local and nonlocal unitary operations can alter the amount of coherence in a quantum system as they transform the original basis. Quantum coherence is an important physical resource and, like entanglement, plays a central role in various quantum information and estimation protocols, and offers advantages over classical ones [4, 5]. In recent years, several crucial attempts [6–8] have been made to characterize and quantify coherence. Baumgratz *et al.* [8] have provided a rigorous axiomatic framework to quantify coherence, in analogy with entanglement, for general quantum systems. In the resource theory of quantum coherence, "incoherent quantum states" are considered free states and "incoherent quantum channels" as free operations that map the set of incoherent states onto itself (see Sec. II). Being a useful resource, several measures of quantum coherence namely the $K$ coherence, robustness of coherence, distance-based coherence, relative entropy of coherence, partial coherence, etc., have been introduced [8–20]. Moreover, quantum coherence and its distribution in multipartite systems have also been studied [21, 22].

Because the superposition principle is the essence of both quantum coherence and quantum correlations, it is interesting and relevant to investigate quantitative relations between them, i.e., how can one resource emerge from the other [23, 24]. Recently, Streltsov *et al.* [25] proposed an interesting framework for the interconversion between coherence and entanglement and thus provided a quantitative connection between them. Furthermore, the interplay between coherence and quantum discord [26–28] has been investigated in Ref. [29]. Also, it has been demonstrated that coherence can be converted into other quantum resources beyond quantum correlations. For example, coherence is a fundamental resource in creation of "magic" [30] and it can be interconverted into resources like quantum Fisher information, superradiance and entanglement under incoherent operations [31]. In this brief report, we follow a mathematical and logical approach to study the bidirectional relationship between partial coherence and quantum correlations (quantified by quantum discord).

The paper is structured as follows. In Sec. II, we review quantum coherence and quantum discord. In Sec. III, We first show that (partial) coherence can be converted to quantum correlations via partial incoherent operations, and conversely, a new relationship from quantum correlations to partial coherence via local unitary operations is presented. In Sec. IV, we propose families of new coherence measures using quantum correlations and quantum Fisher information. Finally, Sec. V concludes with summary.

## II. PARTIAL COHERENCE AND DISCORD

### A. Coherence and Partial Coherence

For a fixed orthonormal basis $\{|i\rangle\}$ (alternatively, a von Neumann measurement $\Pi = \{\Pi_i = |i\rangle\langle i|\}$) of a system $\mathcal{H}$, we review the axiomatic notion of coherence as proposed by Baumgratz *et al* [8].

(i) The set of incoherent states is defined by

$$\mathcal{I} = \big\{\sigma = \sum_i p_i |i\rangle\langle i| : p_i \geq 0, \sum_i p_i = 1\big\}.$$

---
[*] kimshanhao@126.com
[†] Corresponding author: lilongsuo@126.com
[‡] Corresponding author: asutoshk.phys@gmail.com
[§] Corresponding author: wjd@zju.edu.cn

(ii) A completely positive trace preserving (CPTP) map $\Phi$ is said to be an incoherent operation if it can be written as

$$\Phi(\sigma) = \sum_k E_k \sigma E_k^\dagger,$$

where $E_k$'s are "incoherent" Kraus opertors in the sense $E_k \mathcal{I} E_k^\dagger \subseteq \mathcal{I}$. We denote the set of all incoherent operations by $\mathcal{O}$.

A functional $C(\rho|\Pi)$ on the space of quantum states is regarded as a coherence measure (with respect to the von Neumann measurement $\Pi$), if it satisfies the following conditions.

(C1) *Nonnegativity.* $C(\rho|\Pi) \geq 0$, and $C(\rho|\Pi) = 0$ if and only if $\rho \in \mathcal{I}$.

(C2a) *Weak monotonicity.* $C(\rho|\Pi)$ is nonincreasing under incoherent operations, i.e., $C(\rho|\Pi) \geq C(\Phi(\rho)|\Pi)$ with $\Phi(\mathcal{I}) \subseteq \mathcal{I}$.

(C2b) *Strong monotonicity.* $C(\rho|\Pi)$ is nonincreasing on average under selective incoherent operations, i.e., $C(\rho|\Pi) \geq \sum_k p_k C(\varrho_k|\Pi)$, where $p_k = \text{tr}(E_k \rho E_k^\dagger)$ and $\varrho_k = E_k \rho E_k^\dagger / p_k$ for incoherent Kraus operators $E_k$.

(C3) *Convexity.* $C(\rho|\Pi)$ is nonincreasing under mixing of quantum states.

Note that convexity (C3) followed by strong monotonicity (C2b) imply weak monotonicity (C2a). We, therefore, call weak coherence and strong coherence when the quantifier satisfies weak monotonicity (C2a) and strong monotonicity (C2b), respectively. Within such a framework of coherence, one can define suitable measures that satisfy above requirements, and are called coherence monotones. Coherence measures based on the relative entropy and the $l_1$-norm are examples of coherence monotones [28]. For a quantum state $\rho$, in the reference basis $\{|i\rangle\}$, they are defined as following: $\mathcal{C}_{l_1}(\rho) := \sum_{i \neq j} |\rho_{ij}|$ and $\mathcal{C}_r(\rho) := \min_{\sigma \in \mathcal{I}} S(\rho \parallel \sigma) = S(\rho^d) - S(\rho)$, where $\mathcal{I}$ is the set of all incoherent states in the reference basis $\{|i\rangle\}$, $S(\rho \parallel \sigma) = \text{Tr}\rho(\log \rho - \log \sigma)$ is the relative entropy between $\rho$ and $\sigma$, $S(\rho) = -\text{Tr}\rho \log \rho$ is the von Neumann entropy of $\rho$, and $\rho^d$ is the diagonal state of $\rho$, i.e., $\rho^d = \sum_i \langle i|\rho|i\rangle |i\rangle\langle i|$.

Next, let $\rho^{ab}$ be a bipartite quantum state shared by parties $a$ and $b$, and $\Pi_L = \{\Pi_i^a \otimes \mathbf{1}^b\}$ is the Lüders measurement extension of a fixed local von Neumann measurement $\Pi^a = \{\Pi_i^a\}$ on party $a$. Then the notion of "partial coherence" with respect to the Lüders measurement $\Pi_L$ is as follows [18].

(i) The set of partial incoherent states is defined by

$$\mathcal{I}_P^a = \{\sigma^{ab} : \Pi_L(\sigma^{ab}) = \sigma^{ab}\},$$

where $\Pi_L(\sigma^{ab}) = \sum_i (\Pi_i^a \otimes \mathbf{1}^b) \sigma^{ab} (\Pi_i^a \otimes \mathbf{1}^b)$.

(ii) A CPTP map $\Lambda^a$, with Kraus operators $\{K_l\}$, is called partial incoherent if $K_l \mathcal{I}_P^a K_l^\dagger \in \mathcal{I}_P^a$. We denote the set of all partial incoherent operations by $\mathcal{O}_P^a$.

A functional $C^a(\rho^{ab}|\Pi_L)$ on the space of bipartite quantum states $\rho^{ab}$ is a measure of partial coherence (with respect to $\Pi_L$), if it satisfies the following conditions.

(P1) *Nonnegativity.* $C^a(\rho^{ab}|\Pi_L) \geq 0$, and $C^a(\rho^{ab}|\Pi_L) = 0$ if and only if $\rho^{ab} \in \mathcal{I}_P^a$.

(P2a) *Weak monotonicity.* $C^a(\rho^{ab}|\Pi_L)$ is nonincreasing under partial incoherent operations, i.e., $C^a(\rho^{ab}|\Pi_L) \geq C^a(\Lambda^a(\rho^{ab})|\Pi_L)$ with $\Lambda^a(\mathcal{I}_P^a) \subseteq \mathcal{I}_P^a$.

(P2b) *Strong monotonicity.* $C^a(\rho^{ab}|\Pi_L)$ is nonincreasing on average under selective partial incoherent operations, i.e., $C^a(\rho^{ab}|\Pi_L) \geq \sum_l p_l C^a(\varsigma_l|\Pi_L)$, where $p_l = \text{tr}(K_l \rho K_l^\dagger)$ and states $\varsigma_l = K_l \rho K_l^\dagger / p_l$ for partial incoherent Kraus operators $K_l$.

(P3) *Convexity.* $C^a(\rho^{ab}|\Pi_L)$ is convex with respect to mixing of quantum states.

Very recently, Luo and Sun in Ref. [18] have proposed a measure of partial coherence, $C_I^a$, which satisfies the weak monotonicity condition (P2a), and is given by

$$C_I^a(\rho^{ab}|\Pi_L) \equiv \sum_i I(\rho^{ab}, \Pi_i^a \otimes \mathbf{1}^b), \quad (1)$$

where $I(\sigma, K) = -\frac{1}{2} \text{tr}([\sqrt{\sigma}, K]^2)$ is Wigner-Yanase skew information [9]. Unlike coherence which is defined for a single system as well, partial (quantum) coherence is defined in bipartite systems and has a direct relationship to quantum correlations.

### B. Quantum Discord

Quantum discord characterizes "nonclassicality" of correlations in quantum mechanics beyond entanglement in the sense that it may not vanish even for (mixed) separable quantum states, $\rho^{ab} = \sum_i p_i \rho_i^a \otimes \rho_i^b$ [32]. There are two important versions of quantum discord in the quantum information literature: (i) Entropic quantum discord (EQD) [26–28] which is the minimal difference between *quantum mutual information* before and after measurement on a bipartite quantum state, and (ii) Geometric quantum discord (GQD) [33, 34] which captures quantum correlations from a geometric perspective, and coincides with a simpler quantity based on von Neumann measurements. To serve our purpose, we consider GQD only. We review GQD (the modified version) as the minimal partial coherence introduced in Ref. [35]. The GQD based on Wigner-Yanase skew information ($I(\sigma, K) = -\frac{1}{2} \text{tr}([\sqrt{\sigma}, K]^2)$) can be alternatively expressed as [17, 36]

$$Q_G^a(\rho^{ab}) \equiv \min_{\Pi^a} \|\sqrt{\rho^{ab}} - \Pi^a \otimes \mathbb{I}^b(\sqrt{\rho^{ab}})\|^2 \quad (2)$$

$$= \min_{\Pi^a} \sum_i I(\rho^{ab}, \Pi_i^a \otimes \mathbf{1}^b). \quad (3)$$

### III. PARTIAL COHERENCE AND QUANTUM CORRELATIONS

Streltsov *et al.* [37] have shown recently that only the coherent states can be converted to entanglement via incoherent operations. We establish a similar result here for quantified quantum correlations (quantum discord) instead of entanglement. In particular, quantum correlations can be generated by partial incoherent operations if and only if the input state $\rho^{ab}$ is partial coherent (see Fig. 1). To serve our purpose, we consider partial coherence ($C_I^a$ in Eq. (1)) and GQD ($Q_G^a$ in Eq. (3)) based on Wigner-Yanase skew information defined earlier. If the input state is of the product form $\rho^a \otimes \rho^b$, where $\rho^a$ and $\rho^b$ are quantum states on the target system $\mathcal{H}^a$ and ancilla system $\mathcal{H}^b$ respectively, then from the additivity of skew information, the amount of partial coherence for the bipartite state coincides with the amount of coherence for $\rho^a$ on party $a$ as

$$C_I^a(\rho^a \otimes \rho^b | \Pi_L) = \sum_i I(\rho^a \otimes \rho^b, \Pi_i^a \otimes \mathbf{1}^b)$$
$$= \sum_i I(\rho^a, \Pi_i^a) = C_I^a(\rho^a | \Pi^a), \quad (4)$$

where $\Pi^a = \{\Pi_i^a\}$ is the von Neumann measurement on party $a$. This holds true for any ancilla system $\mathcal{H}^b$. The following theorem introduces an upper bound on the amount of quantum correlation that can be generated by applying a partial incoherent operation.

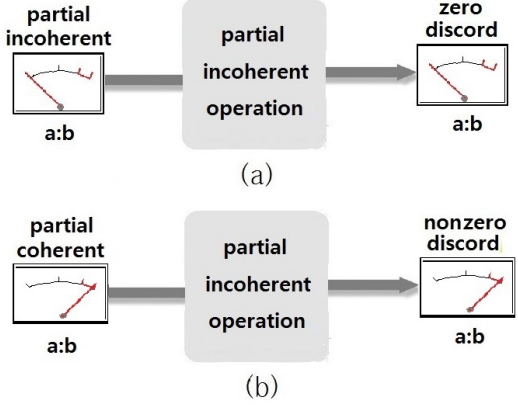

FIG. 1. (a) Partial incoherent operations cannot generate quantum correlations from partial incoherent input states of quantum system $\mathcal{H}^a \otimes \mathcal{H}^b$. (b) On the contrary, if the input state is partial coherent, the partial coherence can be transformed into quantum correlations via partial incoherent operations. However, the amount of quantum correlations generated from partial coherence does not exceed the amount of original partial coherence.

**Theorem 1.** *The quantity of quantum correlation $Q_G^a$ generated from a state $\rho^{ab}$ via a partial incoherent operation $\Lambda^a$ is upper bounded by its partial coherence $C_I^a$, i.e.,*

$$Q_G^a\{\Lambda^a(\rho^{ab})\} \leq C_I^a(\rho^{ab} | \Pi_L).$$

*Furthermore, it induces the following:*

$$Q_G^a\{\Lambda^a(\rho^a \otimes |0\rangle\langle 0|^b)\} \leq C_I(\rho^a | \Pi^a).$$

*Proof.* By definitions of $Q_G^a$ and $C_I^a$, we have $Q_G^a(\rho^{ab}) \leq C_I^a(\rho^{ab} | \Pi_L)$. Also, the inequality

$$Q_G^a\{\Lambda^a(\rho^{ab})\} \leq C_I^a(\Lambda^a(\rho^{ab}) | \Pi_L) \leq C_I^a(\rho^{ab} | \Pi_L)$$

follows from the weak monotonicity of partial coherence measure $C_I^a$. Further, Eq. (4) induces

$$Q_G^a\{\Lambda^a(\rho^a \otimes |0\rangle\langle 0|^b)\} \leq C_I(\rho^a | \Pi^a).$$

□

Clearly, from the above result one can see that quantum correlation and partial coherence, like entanglement and coherence, are closely connected. Another thing to note is that mere the presence of partial coherence does not guarantee the existence of quantum correlations (by definition, the latter is always smaller than the former), but the former can be transformed into the latter by a partial incoherent operation. This shows that partial coherence is a resource. The next theorem tells us which quantum states can have nonzero quantum correlations on application of partial incoherent operations onto them.

**Theorem 2.** *A quantum state $\rho^{ab}$ can be converted to a nonzero geometric discord state via partial incoherent operations if and only if $\rho^{ab}$ is partial coherent.*

*Proof.* By Theorem 1, if $\rho^{ab}$ is partial incoherent, it cannot be converted to a non-zero discord state via partial incoherent operations.

Next, when $\rho^{ab}$ is partial coherent, let $\Pi^a = \{\Pi_i^a = |i\rangle\langle i|^a\}$ is the local von Neumann measurement on party $a$ to determine partial coherence. If a partial incoherent operation $\Lambda^a$ is such that $\Lambda^a(\rho^{ab})$ is a zero discord state

$$\Lambda^a(\rho^{ab}) = \sum_j p_j \sum_l K_l |\psi_j\rangle\langle\psi_j| K_l^\dagger$$
$$= \sum_{n,k} p(n,k) |\phi_n\rangle\langle\phi_n|^a \otimes |\varphi_{n,k}\rangle\langle\varphi_{n,k}|^b,$$

where $\rho^{ab} = \sum_j p_j |\psi_j\rangle\langle\psi_j|$ is a spectral decomposition, it means that $K_l |\psi_j\rangle\langle\psi_j| K_l^\dagger$ are separable for all $j, l$. Namely, for some $n$ and $k$,

$$K_l |\psi_j\rangle = s_{j,l} |\phi_n\rangle^a |\varphi_{n,k}\rangle^b,$$

where $|s_{j,l}|^2 = \text{tr}(K_l |\psi_j\rangle\langle\psi_j| K_l^\dagger)$. Therefore, when $|\psi_j\rangle = \sum_i q_i^{(j)} |i\rangle^a |v_i^{(j)}\rangle^b$ is a decomposition ($\{|v_i\rangle^b\}_i$ need not orthogonal), all $K_l$ must satisfy at least one of the following two conditions for any $j$:

(i) $\quad K_l |i\rangle^a |v_i^{(j)}\rangle^b = a_{i,l}^{(j)} |\tau_{i,l}^{(j)}\rangle^a |\varphi_{n,k}\rangle^b,$

where $\sum_i q_i^{(j)} a_{i,l}^{(j)} |\tau_{i,l}^{(j)}\rangle^a = s_{j,l} |\phi_n\rangle^a$.

(ii) $\quad K_l |i\rangle^a |v_i^{(j)}\rangle^b = b_{i,l}^{(j)} |\phi_n\rangle^a |\varsigma_{i,l}^{(j)}\rangle^b,$

where $\sum_i q_i^{(j)} b_{i,l}^{(j)} |\varsigma_{i,l}^{(j)}\rangle^a = s_{j,l} |\varphi_{n,k}\rangle^b$.

Let $\mathcal{O}_{sep}^a$ be the set of all partial incoherent operations together with the partial incoherent Kraus opertors satisfying at least one of the above two conditions. Then we certainly have the relationship $\mathcal{O}_{sep}^a \subsetneq \mathcal{O}_P^a$. Consequently, for any $\Lambda^a \in (\mathcal{O}_{sep}^a)^c \cap \mathcal{O}_P^a$, $\Lambda^a(\rho^{ab})$ is a non-zero discord state. □

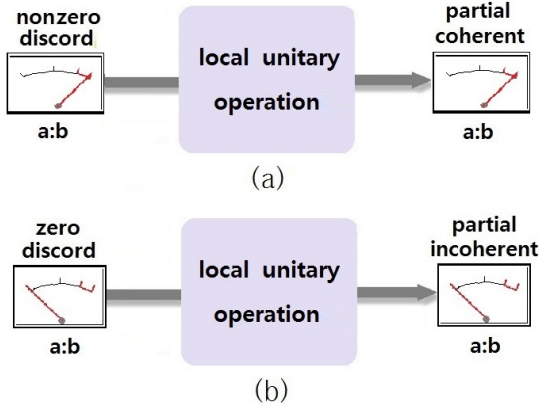

FIG. 2. (a) Nonzero discord states are in themselves partial coherent states and the partial coherence does not disappear via any local unitary operations on parties $a$ and $b$. (b) On the contrary, a zero discord state itself is partial incoherent state or there exists a local unitary operation that causes the partial coherence to vanish.

*Remark.* Previous studies have considered generation of quantum correlations from the separable form of a coherent state (in the single system) and an auxiliary system. However, in the case of partial coherent state, it is possible to create quantum correlations without the need for an auxiliary system. [A partial coherent state does not need another auxiliary system because it is itself a bipartite system. Also, since the bipartite system itself is the original system, it is difficult to identify the part of the partial coherent state on which the measurement is not done as the auxiliary system. Generally, the separable form of a coherent state (in the single system) and an auxiliary system is $\rho_S \otimes \rho_A$, but the partial coherent states are possible even if they are not in this form. However, if we compare it with the coherence defined in a single system, we can say that the part on which the measurement is not done can be regarded as an auxiliary system.] In particular, a quantum state $\rho^a$ can be converted to a nonzero discord state via partial incoherent operations if and only if $\rho^a$ is coherent. This is because if $\rho^a$ is coherent (in the single system), $\rho^a \otimes |0\rangle\langle 0|^b$ becomes partial coherent in the bipartite system (see Eq. (4)).

As shown in Fig. 2, we note two important observations.

(i) When $\rho^{ab}$ is a nonzero discord state, $\mathcal{Q}^a_G$ is invariant under any local unitary operation $\Phi_U$ on parties $a$ and $b$. That is,

$$0 < \mathcal{Q}^a_G(\rho^{ab}) = \mathcal{Q}^a_G\{\Phi_U(\rho^{ab})\} \le C^a_I\{\Phi_U(\rho^{ab})|\Pi_L\}.$$

In other words, this means that $\Phi_U(\rho^{ab})$ is a partial coherent for any local unitary operation $\Phi_U$.

(ii) For a zero discord state $\rho^{ab} = \sum_n p(n)|\phi_n\rangle\langle\phi_n|^a \otimes \sigma^{b|n}$ (especially, when $\{|\phi\rangle_n\}_n$ is not equal to $\{|i\rangle\}_i$), if $\Phi_U$ is a local unitary operation (including the local unitary Kraus operators to exchange $\{|\phi_n\rangle\}_n$ for $\{|i\rangle\}_i$) on party $a$, then we can easily see that $\Phi_U(\rho^{ab})$ is a partial incoherent state.

This means that there is not only a one-sided relationship from partial coherence to quantum correlations, but also exists a close relationship in both directions. The bipartite quantum states having nonvanishing quantum correlations (converted from partial coherent states via partial incoherent operations) still maintain partial coherence, and the amount of partial coherence is not less than the amount of quantum correlations obtained via any local unitary operation. Only the partial coherence inherent in zero discord states is completely lost via local unitary operations.

## IV. COHERENCE VIA QUANTUM CORRELATIONS AND QUANTUM FISHER INFORMATION

When $Q^a$ is an arbitrary convex measure of quantum correlations and the supremum is taken over all partial incoherent operations $\Lambda^a$, we define a family of coherence measures based on quantum correlations as follows.

1. Weak partial coherence via quantum correlations

$$C^a_{Q,w}(\rho^{ab}|\Pi_L) = \sup_{\Lambda^a} Q^a\{\Lambda^a(\rho^{ab})\},$$

2. Weak coherence via quantum correlations

$$C_{Q,w}(\rho^a|\Pi^a) = \sup_{\Lambda^a} Q^a\{\Lambda^a(\rho^a \otimes |0\rangle\langle 0|^b))\}.$$

3. Strong coherence via quantum correlations

$$C_{Q,s}(\rho^a|\Pi^a) = \lim_{N_b \to \infty}\left[\sup_{\Lambda^a} Q^a\{\Lambda^a(\rho^a \otimes |0\rangle\langle 0|^b)\}\right],$$

where $N_b$ is the dimension of the ancilla system $\mathcal{H}^b$. Here we assume that $Q^a$ satisfies the strong monotonicity condition

$$\begin{aligned}Q^a\Big(\sum_i p_i \rho^{ab}_i \otimes |i\rangle\langle i|^c\Big) &= \sum_i p_i Q^a\Big(\rho^{ab}_i \otimes |i\rangle\langle i|^c\Big) \\ &\ge \sum_i p_i Q^a(\rho^{ab}_i),\end{aligned} \quad (5)$$

where $c$ is another ancilla particle and $|i\rangle\langle i|^c$ are orthonormal to each other. Measures of quantum correlations quantified by Wigner-Yanase skew information and quantum Fisher information, in general, satisfy this condition [35, 36, 38, 39].

We first show that $C^a_{Q,w}$ satisfies conditions P1, P2a and P3 to prove that the measure is properly defined.

**Theorem 3.** $C^a_{Q,w}$ *is a weak partial coherence measure for any convex measure of quantum correlation $Q^a$.*

*Likewise, $C_{Q,w}$ is a weak coherence measure for any convex measure of quantum correlation $Q^a$.*

*Proof.* The proof is similar for both $C^a_{Q,w}$ and $C_{Q,w}$, and therefore we consider the proof of measure $C^a_{Q,w}$ only.

P1. The nonnegativity of $C^a_{Q,w}$ follows from the nonnegative property of $Q^a$, and Theorem 2 induces that $C^a_{Q,w}$ is zero if and only if $\rho^{ab}$ is incoherent.

P2a. For any partial incoherent operation $\Xi^a$,

$$C^a_{Q,w}\{\Xi^a(\rho^{ab})|\Pi_L\} = \sup_{\Lambda^a} Q^a\{\Lambda^a \circ \Xi^a(\rho^{ab})\}.$$

Note that $\Lambda^a \circ \Xi^a$ is also a partial incoherent operation, and $\mathcal{O}_P^a(\Xi^a) = \{\Lambda^a \circ \Xi^a | \Lambda^a \in \mathcal{O}_P^a\} \subset \mathcal{O}_P^a$. Therefore, we have

$$C_{Q,w}^a\{\Xi^a(\rho^{ab})|\Pi_L\} = \sup_{\Lambda^a} Q^a\{\Lambda^a \circ \Xi^a(\rho^{ab})\}$$
$$\leq \sup_{\Lambda^a} Q^a\{\Lambda^a(\rho^{ab})\}$$
$$= C_{Q,w}^a(\rho^{ab}|\Pi_L).$$

P3. The convexity of $Q^a$ yields

$$Q^a\left\{\Lambda^a\left(\sum_i p_i \rho_i^{ab}\right)\right\} \leq \sum_i p_i Q^a\{\Lambda^a(\rho_i^{ab})\},$$

where $\rho^{ab} = \sum_i p_i \rho_i^{ab}$. Taking the supremum over all partial incoherent operations $\Lambda^a$ on both sides of this inequality, we obtain

$$C_{Q,w}^a(\sum_i p_i \rho_i^{ab}|\Pi_L) \leq \sup_{\Lambda^a}\left[\sum_i p_i Q^a\{\Lambda^a(\rho_i^{ab})\}\right]. \quad (6)$$

Also, the RHS of above inequality cannot increase if the supremum is performed on each term of the sum as follows,

$$\sup_{\Lambda^a}\left[\sum_i p_i Q^a\{\Lambda^a(\rho_i^{ab})\}\right] \leq \sum_i p_i \sup_{\Lambda^a} Q^a\{\Lambda^a(\rho_i^{ab})\}$$
$$= \sum_i p_i C_{Q,w}^a(\rho_i^{ab}|\Pi_L). \quad (7)$$

From Ineqs. (6) and (7), the convexity of $C_{Q,w}^a$ is proved. $\square$

Next, we show that $C_{Q,s}$ is a proper coherence measure (in the strong sense) by using a similar line of proof as that for entanglement in Ref. [37].

**Theorem 4.** *$C_{Q,s}$ is a strong coherence measure for any convex measure of quantum correlation $Q^a$ that satisfies condition in Eq. (5).*

*Proof.* We need to prove that $C_{Q,s}$ satisfies the conditions C1, C2b and C3, noting that conditions C2b and C3 together imply C2a. Also, proofs of conditions C1 and C3 are similar to those of P1 and P3 of Theorem 3. Therefore, we prove condition C2b only.

Note that incoherent operations $\Lambda^{ab}$ and $\Lambda^{abc}$ on the systems $ab$ and $abc$ are also partial incoherent operations with respect to $a:b$ and $a:bc$, respectively. Then, by Ineq. (5), we have (see Appendix 1 of Ref. [37])

$$\sum_i p_i C_{Q,s}(\sigma_i^a|\Pi_L) \leq C_{Q,s}(\rho^a|\Pi_L)$$

with probabilities $p_i = \text{tr}(K_i \rho^a K_i^\dagger)$ and quantum states $\sigma_i^a = K_i \rho^a K_i^\dagger / p_i$, where $K_i$'s are partial incoherent Kraus operators. $\square$

We have shown how to quantify coherence via quantum correlations. The problem, however, is that the convexity of quantum correlation measures is rarely known. Therefore, we present a new coherence measure via quantum Fisher information, which ensures convexity in a similar way as above. A version of coherence measure via the quantum Fisher information has been introduced in Ref. [19] for any state $\rho$ on $\mathcal{H}$ as

$$C_F(\rho|\Pi) \equiv \sum_i F(\rho, \Pi_i),$$

where $\Pi$ is von Neuamann measurement determined by coherence, and $F$ is quantum Fisher information

$$F(\sigma, H) = \frac{1}{4}\text{tr}(\sigma L^2), \quad \frac{1}{2}(L\sigma + \sigma L) = i[\sigma, H],$$

defined via logarithmic symmetric derivative. This measure has been established for the following two conditions only:

(F1) *Nonnegativity.* $C_F(\rho|\Pi) \geq 0$, and $C_F(\rho|\Pi) = 0$ is and only if $\rho \in \mathcal{I}$.

(F2) *Convexity.* $C_F(\rho|\Pi)$ is convex in $\rho$.

However, we conjecture another property of $C_F$, namely monotonicity under incoherent operations.

(F3) *Monotonicity.* $C_F\{\Phi(\rho)|\Pi\} \leq C_F(\rho|\Pi)$ for any incoherent operation $\Phi$. To comply this, we define another versions of coherence measure as follows:

1. Weak coherence via quantum Fisher information

$$C_{F,w}(\rho|\Pi) = \sup_{\Phi} C_F\{\Phi(\rho)|\Pi\},$$

for any state $\rho$ on $\mathcal{H}$, and the supremum is taken over all incoherent operations $\Phi$.

2. Strong coherence via quantum Fisher information

$$C_{F,s}(\rho^a|\Pi^a) = \lim_{N_b \to \infty}\left[\sup_{\Lambda^a} C_F\{\Lambda^a(\rho^a \otimes |0\rangle\langle 0|^b)|\Pi_L\}\right]$$

for any state $\rho^a$ on $\mathcal{H}^a$, and the supremum is taken over all partial incoherent operations $\Lambda^a$.

Following the similar lines of proof as in Theorems 3 and 4, one can show explicitly that $C_{F,w}$ and $C_{F,s}$ satisfy all the conditions (nonnegativity, convexity and monotonicity) of a coherence measure.

## V. CONCLUSIONS

In this paper, we investigated the bidirectional relationship between partial coherence and quantum correlations. We showed that the presence of nonzero partial coherence is indispensable (necessary and sufficient) to generate quantum correlations via partial incoherent operations in a bipartite system. In particular, the resources requisite to generate quantum correlations are not limited to the separable form, in which initially a coherent state and an uncorrelated incoherent ancilla are coupled, and it is possible to convert partial coherence of the bipartite system into quantum correlations itself (even if it is of nonseparable form). We found that partial coherence inherent in nonzero discord states is not completely lost via any local unitary operations on parties $a$ and $b$. Also, the amount of discord lower bounds the amount of partial coherence that changes through local unitary operations. Finally, we propose families of coherence measures using quantum correlations and quantum Fisher information.

## ACKNOWLEDGMENTS

This project is supported by the National Natural Science Foundation of China (Grants No. 11171301 and No. 11571307).